\documentclass[pra,twocolumn]{revtex4-1}

\usepackage{amsfonts,amsmath,amssymb,graphicx,natbib,bbm,bm}

\newcommand{\Exp}[1]{e^{#1}}
\newcommand{\wb}{\overline{\omega}}
\newcommand{\wt}{\widetilde{\omega}}
\newcommand{\ext}{\mathrm{ext}}
\newcommand{\phib}{\overline{\phi}}
\newcommand{\phit}{\widetilde{\phi}}
\newcommand{\eb}{\overline{\eta}}

\newcommand{\nb}{\overline{\nu}}
\newcommand{\nt}{\widetilde{\nu}}
\newcommand{\gb}{\bar{g}}
\newcommand{\gt}{\widetilde{g}}

\newcommand{\bra}[1]{\langle #1|}
\newcommand{\ket}[1]{| #1\rangle}
\newcommand{\eff}{\mathrm{eff}}
\newcommand{\BJ}{\mathrm{J}}

\newcommand{\moy}[1]{\langle #1\rangle}
\renewcommand{\a}{\hat{a}}
\newcommand{\A}{\a^\dag}
\newcommand{\hs}{\hat{\sigma}}
\newcommand{\hS}{\hat{s}}

\newcommand{\eigen}{\mathrm{eigen}}
\newcommand{\hU}{\hat{U}}

\newcommand{\ID}{\mathbbm{1}}

\begin{document}

\title{Analytical modeling of parametrically-modulated transmon qubits}

\author{Nicolas Didier}
\affiliation{Rigetti Computing, 775 Heinz Avenue, Berkeley, CA 94710}

\author{Eyob A. Sete}
\affiliation{Rigetti Computing, 775 Heinz Avenue, Berkeley, CA 94710}

\author{Marcus P. da Silva}
\affiliation{Rigetti Computing, 775 Heinz Avenue, Berkeley, CA 94710}

\author{Chad Rigetti}
\affiliation{Rigetti Computing, 775 Heinz Avenue, Berkeley, CA 94710}

\begin{abstract}
Building a scalable quantum computer requires developing appropriate models to understand and verify its complex quantum dynamics. We focus on superconducting quantum processors based on transmons for which full numerical simulations are already challenging at the level of qubytes. It is thus highly desirable to develop accurate methods of modeling qubit networks that do not rely solely on numerical computations. Using systematic perturbation theory to large orders in the transmon regime, we derive precise analytic expressions of the transmon parameters. We apply our results to the case of parametrically-modulated transmons to study recently-implemented parametrically-activated entangling gates.
\end{abstract}

\maketitle

\section{Introduction}

Understanding the complex dynamics of a quantum machine requires the accurate modeling of the individual building blocks: interacting sets of qubits.
Precise understanding is crucial to design, manipulate, optimize, and verify the machine.
In the field of superconducting quantum computers, the transmon~\cite{Koch_2007,xmon} is currently widely used as qubit~\cite{
Kelly15,Barends_2016,Roushan_2017,OMalley_2016,
Kandala_2017,Riste_2017,Takita_2016,IBM16Q,
Langford_2016,Versluis_2016,
Chao_2017,
Salathe_2015,
Hacohen_2016,
Wendin_2016,
Rigetti-blue_2017,Rigetti-white_2017,Rigetti-clustering_2017}
or more general-purpose quantum devices~\cite{Naik_2017,Ofek_2016,Bretheau_2015}.
Transmons, weakly non-linear oscillators derived from a Cooper-pair box, are made from a Josephson tunnel junction shunted by a capacitance. 
The transmon regime corresponds to a large Josephson energy compared to the charging energy --- it is a compromise between a large anharmonicity and a weak sensitivity to charge noise.
The coherence and gate times of transmons in quantum computing experiments have been steadily improving over
the last several years, and transmons are now one of the leading candidates to an architecture that can meet
the stringent requirements of fault-tolerant quantum computing~\cite{Aliferis_2006,Fowler12}.

Although analytical expressions for the behavior of non-interacting transmons are well understood,
the accurate description for the behavior of interacting transmons requires the diagonalization of
coupled systems (i.e., the charge basis description of the transmons 
with charge dipole interactions). Numerical diagonalization of these systems quickly becomes intractable because 
a large number of basis states are necessary to obtain high accuracy even for non-interacting transmons.
A more efficient approach is to use analytical expressions of transmon energies and states.
Exact diagonalization of the Cooper-pair box Hamiltonian is achieved with Mathieu functions~\cite{Cottet_2002,Blais_2004,Koch_2007}, but
manipulating them can be cumbersome.
An alternative is to consider controlled approximations, such as the approximate diagonalization via standard perturbation theory, 
which is widely used in quantum mechanics~\cite{Messiah}. For transmons, the natural small parameter is the ratio of the charging energy of
the Cooper-pair box to the Josephson energy of the junction, as this parameter is typically below $2\,\%$.

In this paper, we revisit transmon qubit theory by utilizing a systematic perturbation theory to model interacting transmons at sub-kHz accuracy with respect to numerical diagonalization.
The resulting analytical expressions are particularly useful to estimate crosstalk in the dispersive regime.
We then use these analytic expressions to model the parametric control of transmon qubits to realize two-qubit gates. 
In our proposal for entangling gates a fixed-frequency transmon is capacitively coupled to a tunable transmon and its flux bias is modulated so as to compensate for the detuning between the two two-qubit transitions of interest. Our gates are thus similar to other proposals for parametric gates~\cite{Bertet,Niskanen2006,Niskanen2007,BeaudoinPRA12,StrandPRB13,Royer_2017,Vijay,McKayPRApp16,Naik_2017}, but distinct in the sense that our proposal directly modulates the qubit used for computation instead of using couplers to mediate the interaction.
Our theory has been already used to predict parametric iSWAP and controlled Z gates that have been successfully realized on 2-qubit~\cite{Rigetti-blue_2017}, 8-qubit~\cite{Rigetti-white_2017}, and  19-qubit processors~\cite{Rigetti-clustering_2017}.

This paper is organized as follows.
We start by presenting the perturbation theory for a single transmon qubit in Sec.~\ref{secF}.
We then consider the case of tunable transmons in Sec.~\ref{secT}.
We treat the capacitive coupling of transmons in Sec.~\ref{secC}.
We use the results to study the physics of parametrically-activated entangling gates in Sec.~\ref{secM}.

\begin{figure}[b]
\includegraphics[width=\columnwidth]{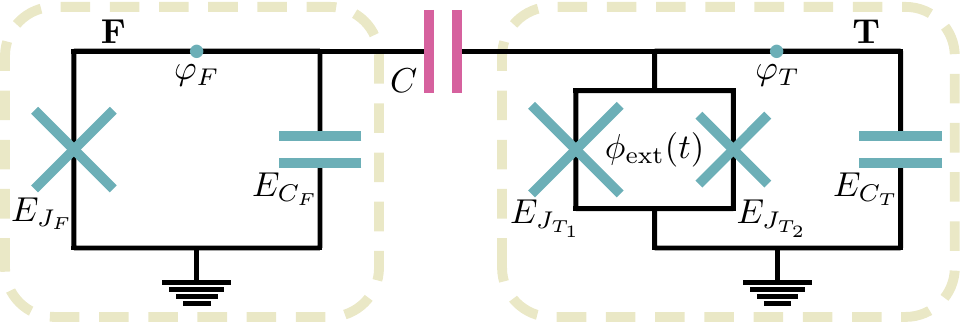}
\caption{Circuit of a fixed-frequency transmon~F (left) and a tunable transmon~T (right) that are capacitively coupled (capacitance $C$).
Transmons are characterized by the charging energy of their capacitance, $E_C=e^2/(2C)$, and the Josephson energy, $E_J$.
The transmon regime of the Cooper pair box corresponds to $E_C/E_J\ll1$. 
Tunable transmons are composed of a SQUID and controlled with pulses on the flux bias line, $\phi_\ext(t)$.}
\label{figtransmon}
\end{figure}

\section{Fixed-frequency transmons}
\label{secF}

The circuit of a fixed-frequency transmon consists of a Josephson junction shunted by a capacitance, as depicted in Fig.~\ref{figtransmon} (left), and is governed by the Hamiltonian,
\begin{align}
\hat{H}_F=4E_C\hat{N}^2-E_J\cos\hat{\varphi}.
\label{Hfixed}
\end{align}
The two conjugate quantum variables here are the Cooper pair number operator $\hat{N}$ and the superconducting phase difference $\hat{\varphi}$ satisfying the commutation rule $[\hat{\varphi},\hat{N}]=i$.
The Schr\"odinger equation for the transmon in the phase representation,
\begin{align}
\left[-4E_C\frac{\partial^2}{\partial\varphi^2}-E_J\cos\varphi\right]\psi(\varphi)=E\psi(\varphi),
\label{SEphase}
\end{align}
can be solved exactly in terms of Mathieu functions~\cite{Cottet_2002,Blais_2004,Koch_2007},
\begin{align}
E_n=&E_C\mathcal{M}_A\!\left(\mu_{n},-\frac{E_J}{2E_C}\right),\label{MathieuE}\\
\psi_n(\varphi)=&\frac{1}{\sqrt{2\pi}}\bigg[\mathcal{M}_C\!\left(\frac{E_n}{E_C},-\frac{E_J}{2E_C},\frac{\varphi}{2}\right)\nonumber\\
&-i^{2n+1}\mathcal{M}_S\!\left(\frac{E_n}{E_C},-\frac{E_J}{2E_C},\frac{\varphi}{2}\right)\bigg],
\label{MathieuPsi}
\end{align}
with 
$\mathcal{M}_A$ the Mathieu characteristic value,
$\mathcal{M}_C$ the even Mathieu function,
$\mathcal{M}_S$ the odd Mathieu function, and
$\mu_{n}= (-1)^{n+1} n + [n\ \mathrm{mod}\ 2]$ the indexes for the eigenenergies.
The structure of the solution indicates that transmons are described with the dimensionless parameter $E_J/(2E_C)$ and the characteristic energy $E_C$.
The transmon wavefunction in Eq.~\eqref{MathieuPsi} is expressed in the phase representation.
Going to the charge representation requires Fourier transforming the Mathieu functions, a calculation that turns out to be rather cumbersome.

The commutation relation between $\hat{\varphi}$ and $\hat{N}$ allows us to express these conjugate variables as the two quadratures of a bosonic field $\a$ (characterized by $[\a,\A]=1$),
\begin{align}
\hat{\varphi}&=\sqrt{\xi}(\A+\a),&
\hat{N}&=\frac{i}{2\sqrt{\xi}}(\A-\a).
\end{align}
The positive real number $\xi$ is related to the zero point fluctuations, $\sqrt{\moy{\hat{\varphi}^2}}=\sqrt{2\xi}$ and $\sqrt{\moy{\hat{N}^2}}=1/\sqrt{2\xi}$.
We then express Hamiltonian Eq.~\eqref{Hfixed} in terms of $\a,\A$
and diagonalize the quadratic part;
It is characterized by the plasma frequency $\omega_h=\sqrt{8E_CE_J}$ and 
solved by assigning the dimensionless parameter $\xi$ to
\begin{align}
\xi=\sqrt{\frac{2E_C}{E_J}}.
\end{align}
In what follows, $\xi$ will be the small parameter of the perturbation theory (typically, $\xi<0.2$).
The transmon Hamiltonian is then expressed as a function of the bosonic field, normal ordered and written as a Taylor series in $\xi$,
\begin{align}
\hat{H}_F&=\sum_{u=0}^\infty\xi^u\hat{H}^{(u)},
\end{align}
with $\hat{H}^{(0)}=\omega_h\,\A\a$ and, for $u\geq1$,
\begin{multline}
\hat{H}^{(u)}=\omega_h\sum_{v=0}^{u}\frac{(-1)^u}{2^{u-v+1}(u-v)!}\\\times\sum_{w=-(v+1)}^{v+1}\frac{\hat{a}^{\dag(v+1+w)}}{(v+1+w)!}\frac{\hat{a}^{(v+1-w)}}{(v+1-w)!},
\end{multline}
expressed in units of $\omega_h$.

\begin{figure}
\includegraphics[width=\columnwidth]{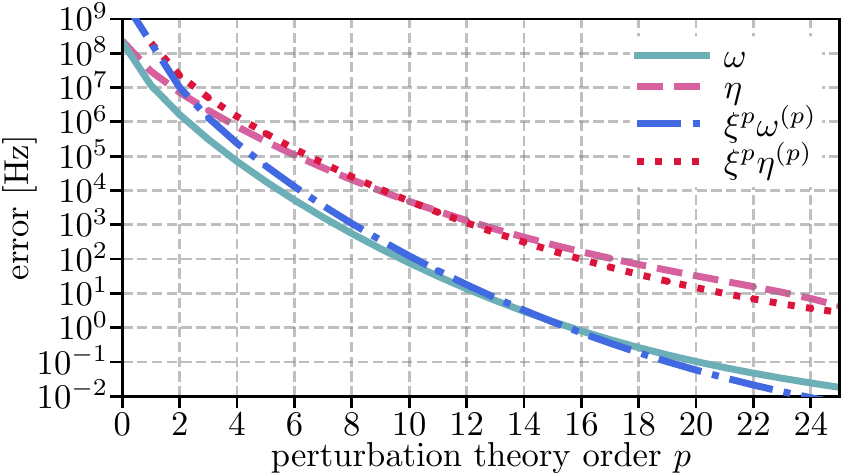}
\caption{Error on the transmon frequency and anharmonicity as a function of perturbation theory order in $\xi$ compared to numerical simulation with 30 Fock states (absolute value of the difference in full lines).
The corresponding terms of the perturbative expansion are plotted in dashed lines.
Sub-kHz accuracy of the transmon frequency and anharmonicity is found at large orders, expressions are reported in Appendix~\ref{App_25}.
Parameters are $\xi=0.2$ and $E_C/h=200\,\mathrm{MHz}$, giving $\omega/(2\pi)=3788\,\mathrm{MHz}$ and $\eta/(2\pi)=230\,\mathrm{MHz}$.}
\label{fig_fixed}
\end{figure}

The eigenenergies $E_n$ and eigenstates $\ket{\psi_n}$ are then obtained using perturbation theory~\cite{Messiah}
in $\xi$,
\begin{align}
E_n&=\sum_{p=0}^\infty\xi^pE_n^{(p)},&
\ket{\psi_n}&=\sum_{p=0}^\infty\xi^p\ket{\psi_n^{(p)}},
\end{align}
by solving the Schr\"odinger equation at each order in $\xi$.
The unperturbed system is a harmonic oscillator of frequency $\omega_h$, the energies are $E_n^{(0)}=n\omega_h$ and the corresponding eigenstates are the Fock states $\ket{\psi_n^{(0)}}=\ket{n}$.
The eigenenergies and eigenstates at order $\xi^p$, for $p\geq1$, are obtained by recurrence,
\begin{align}
E_n^{(p)}=&\sum_{q=0}^{p-1}\bra{n}\hat{H}^{(p-q)}\ket{\psi_n^{(q)}},\label{eigenenergies}\\
\ket{\psi_n^{(p)}}=&\sum_{m\neq n}\bigg\{\frac{1}{(n-m)\omega_h}\bigg[\bra{m}\hat{H}^{(p)}\ket{n}\nonumber\\
+&\sum_{q=1}^{p-1}\bra{m}\hat{H}^{(p-q)}-E_n^{(p-q)}\ket{\psi_n^{(q)}}\bigg]\bigg\}\ket{m}.
\label{eigenstates}
\end{align}
To compute the eigenenergies and eigenstates of level $n$ at order $p$, Fock states $\ket{0}\to\ket{n+4p}$ are required because terms such as $\a^{\dag2(q+1)}$ and $\a^{2(q+1)}$ are involved.
In particular at each new order the Hilbert space is extended by the action of $\a^{\dag4}$ and $\a^{4}$.
The eigenstates $\ket{\psi_n}$ derived by recurrence are not normalized.
The normalized states $\ket{\Psi_n}$ are easily obtained via
$\ket{\Psi_n}=\ket{\psi_n}/\sqrt{\langle\psi_n\ket{\psi_n}}$.
The diagonalization transformation is represented by the operator $\hU_\eigen$ that transforms the $n+4p$ Fock states into the $n$ eigenstates.

The expression of the diagonalization operator for the first 5 transmon eigenstates at $5^\mathrm{th}$~order in $\xi$ is provided in Appendix~\ref{App_Ueigen}. 
In the following we consider transmons as three-level systems, characterized by their frequency $\omega=E_1-E_0$ and anharmonicity $\eta=(E_1-E_0)-(E_2-E_1)$, positive by definition.
At $5^\mathrm{th}$~order, the transmon frequency and anharmonicity read
\begin{align}
\omega\simeq&\sqrt{8E_CE_J}\nonumber\\
-&E_C\bigg[1
+\frac{1}{2^2}\xi
+\frac{21}{2^7}\xi^2
+\frac{19}{2^7}\xi^3
+\frac{5319}{2^{15}}\xi^4
\bigg],\label{perturbationexpandwq}\\
\eta\simeq&E_C\bigg[1
+\frac{9}{2^4}\xi
+\frac{81}{2^7}\xi^2
+\frac{3645}{2^{12}}\xi^3
+\frac{46899}{2^{15}}\xi^4
\bigg].
\label{perturbationexpandeta}
\end{align}
The first five orders are identical to the available asymptotic expansion of Mathieu functions for large arguments~\cite{Abramowitz_book}.
Higher order expansions are easily obtained through these recursive expressions, and as an example, we report $25^\mathrm{th}$~order expansions in Appendix~\ref{App_25}.
The accuracy of the perturbation theory with respect to the numerical diagonalization of the Hamiltonian of a transmon qubit is plotted in Fig.~\ref{fig_fixed}.

\section{Frequency-tunable transmons}
\label{secT}

\begin{figure}[t]
\includegraphics[width=\columnwidth]{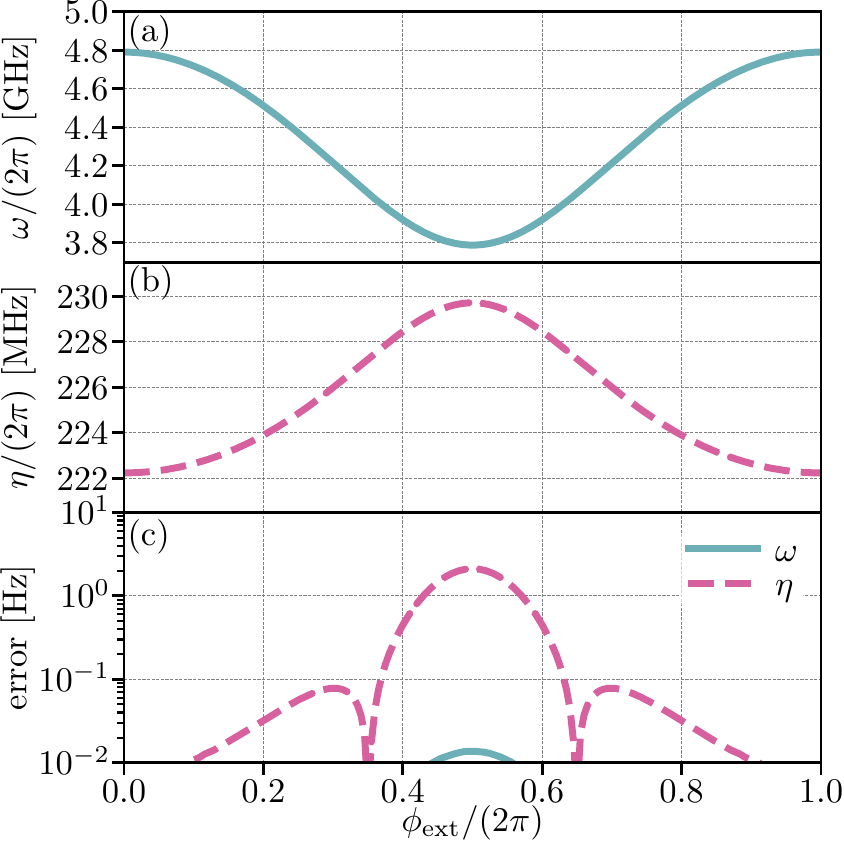}
\caption{Error on the tunable transmon frequency and anharmonicity at $25^\mathrm{th}$ order in $\xi$ as a function of parking flux bias compared to numerical simulation with 30 Fock states.
Parameters are $E_C/(2\pi)=200\,\mathrm{MHz}$, $\xi_\mathrm{max}=0.16$ and $\xi_\mathrm{min}=0.2$, giving $\omega_\mathrm{max}/(2\pi)=4791\,\mathrm{MHz}$, $\omega_\mathrm{min}/(2\pi)=3788\,\mathrm{MHz}$, $\eta_\mathrm{max}/(2\pi)=222\,\mathrm{MHz}$ and $\eta_\mathrm{min}/(2\pi)=230\,\mathrm{MHz}$ 
(subscripts max, min refer to $\phi_\ext=0,\pi$ respectively).}
\label{fig_tunable}
\end{figure}

The transmon frequency can be tuned with an external magnetic field $\phi_\ext$ by replacing the Josephson junction with a SQUID, see Fig.~\ref{figtransmon} (right).
The tunable-transmon Hamiltonian is then that of a split Cooper-pair box,
\begin{align}
\hat{H}_T&=4E_C\hat{N}^2-E_{J_1}\cos(\hat{\varphi}-\phi_\ext)-E_{J_2}\cos\hat{\varphi},
\end{align}
where $E_{J_1}$ and $E_{J_2}$ are the Josephson energies of the SQUID loop.
The Hamiltonian can be recast as an effective single-junction transmon,
\begin{align}
\hat{H}_T&=4E_C\hat{N}^2-E_{J_\eff}\cos(\hat{\varphi}-\phi_\eff),
\end{align}
with the flux-dependent effective Josephson energy and offset phase,
\begin{align}
E_{J_\eff}&=\sqrt{E_{J_1}^2+E_{J_2}^2+2E_{J_1}E_{J_2}\cos\phi_\ext},\\
\phi_\eff&=\arctan\!\left(\frac{\sin\phi_\ext}{\cos\phi_\ext+\frac{E_{J_2}}{E_{J_1}}}\right).
\end{align}
The expression of the effective Josephson energy reveals ``sweet spots'' for flux noise, points of operation that are insensitive at first order to fluctuations of flux bias.
Sweet spots are located at the extrema of the energy spectrum, e.g.~at $\phi_\ext=0,\pi$.
The phase is localized around the offset phase~$\phi_\eff$;
it can be removed with the displacement operator
\begin{align}
\hU_\phi&=\Exp{i\phi_\eff\hat{N}},
\end{align}
giving the same form as the fixed-frequency Hamiltonian Eq.~\eqref{Hfixed},
$\hat{H}_T=4E_C\hat{N}^2-E_{J_\eff}\cos\hat{\varphi}$.
The charge operator is invariant under this unitary transformation --- $U_\phi$ will not affect, e.g., the capacitive coupling interaction.
The perturbation theory developed for a fixed-frequency transmon can be applied to the effective Hamiltonian with the small parameter 
$\xi=\sqrt{2E_C/E_{J_\eff}}$.
The accuracy of the perturbation theory with respect to the numerical diagonalization of the Hamiltonian of a tunable transmon qubit is plotted in Fig.~\ref{fig_tunable}.

For time-dependent flux biases, $\phi_\ext(t)$, the unitary transformation for phase displacement $\hU_\phi(t)$ and diagonalization $\hU_\eigen$ generates nonadiabatic terms,
\begin{align}
&\hat{H}_T(t)=\hU_\eigen\hU_\phi\hat{H}\hU_\phi^\dag\hU_\eigen^\dag\nonumber\\
&\qquad\qquad+i\hU_\eigen\dot{\hU}_\phi\hU_\phi^\dag\hU_\eigen^\dag+i\dot{\hU}_\eigen\hU_\eigen^\dag\\
&=\sum_{n=0}^2E_{n}\hat{\Pi}_n
-\frac{\dot{\phi}_\eff(t)}{2\sqrt{\xi(t)}}[\lambda(t)\hat{\sigma}_{y}
+\sqrt{2}\Lambda(t)\hat{s}_{y}]
-\frac{\dot{\xi}(t)}{\xi(t)}\upsilon(t)\hat{S}_{y},
\end{align}
where we note $\hat{\Pi}_n=\ket{\Psi_{n}}\bra{\Psi_{n}}$ the projector on transmon eigenstate $\ket{\Psi_{n}}$.
The nonadiabatic Hamiltonian acts as a drive between the different transmon levels.
The ladder operators between the three transitions are
\begin{align}
\hat{\sigma}_+&=\ket{\psi_1}\bra{\psi_0},&
\hat{\sigma}_-&=\hat{\sigma}_+^\dagger,&
\hat{\sigma}_y&=i(\hat{\sigma}_+-\hat{\sigma}_-),\\
\hat{s}_+&=\ket{\psi_2}\bra{\psi_1},&
\hat{s}_-&=\hat{s}_+^\dagger,&
\hat{s}_y&=i(\hat{s}_+-\hat{s}_-),\\
\hat{S}_+&=\ket{\psi_2}\bra{\psi_0},&
\hat{S}_-&=\hat{S}_+^\dagger,&
\hat{S}_y&=i(\hat{S}_+-\hat{S}_-).
\end{align}
The parameters $\lambda$, $\Lambda$ are the weights of the charge number operator in the three-level transmon eigenbasis,
\begin{align}
\hat{N}&=\frac{\lambda}{2\sqrt{\xi}}\hs_y+\frac{\Lambda}{\sqrt{2\xi}}\hS_y,
\label{Ntilde}
\end{align}
and $\upsilon$ comes from $\dot{\hU}_\eigen \hU_\eigen^\dag$. 
Keeping the first five terms in the expansion, we get,
\begin{align}
\lambda&=
1 -\frac{1}{2^{3}}\xi -\frac{11}{2^8}\xi^2 -\frac{65}{2^{11}}\xi^3 -\frac{4203}{2^{17}}\xi^4,\\
\Lambda&=
1 -\frac{1}{2^2}\xi -\frac{73}{2^{9}}\xi^2 -\frac{79}{2^9}\xi^3 -\frac{113685}{2^{19}}\xi^4,\\
\upsilon&=-\sqrt{2}\left[\frac{1}{2^4}\xi+\frac{11}{2^{8}}\xi^2+\frac{321}{2^{13}}\xi^3+\frac{5609}{2^{17}}\xi^4\right].
\end{align}
Expressions at higher orders are given in Appendix~\ref{App_25}.

\section{Capacitively-coupled transmons}
\label{secC}

The capacitive coupling of two Cooper-pair boxes, as depicted in Fig.~\ref{figtransmon}, generates a charge-charge interaction through the coupling capacitance, $g_C\hat{N}_1\hat{N}_2$.
From Fig.~\ref{figtransmon}, $g_C=\frac{(2e)^2C}{C_FC_T}$.
To treat the capacitive coupling in the transmon eigenbasis, we use Eq.~\eqref{Ntilde} 
to obtain the coupling Hamiltonian,
\begin{align}
\hat{H}_C&=g_{11}\hat{\sigma}_{y_1}\hat{\sigma}_{y_2}+g_{12}\hat{\sigma}_{y_1}\hat{s}_{y_2}+g_{21}\hat{s}_{y_1}\hat{\sigma}_{y_2}+g_{22}\hat{s}_{y_1}\hat{s}_{y_2},
\label{H_coupled}
\end{align}
and define the coupling strengths,
\begin{align}
g&=\frac{g_C}{4\sqrt{\xi_1\xi_2}},&
g_{11}&=g\lambda_1\lambda_2,&
g_{21}&=\sqrt{2}g\Lambda_1\lambda_2,\nonumber\\&&
g_{22}&=2g\Lambda_1\Lambda_2,&
g_{12}&=\sqrt{2}g\lambda_1\Lambda_2.
\end{align}

\begin{figure}[t]
\includegraphics[width=\columnwidth]{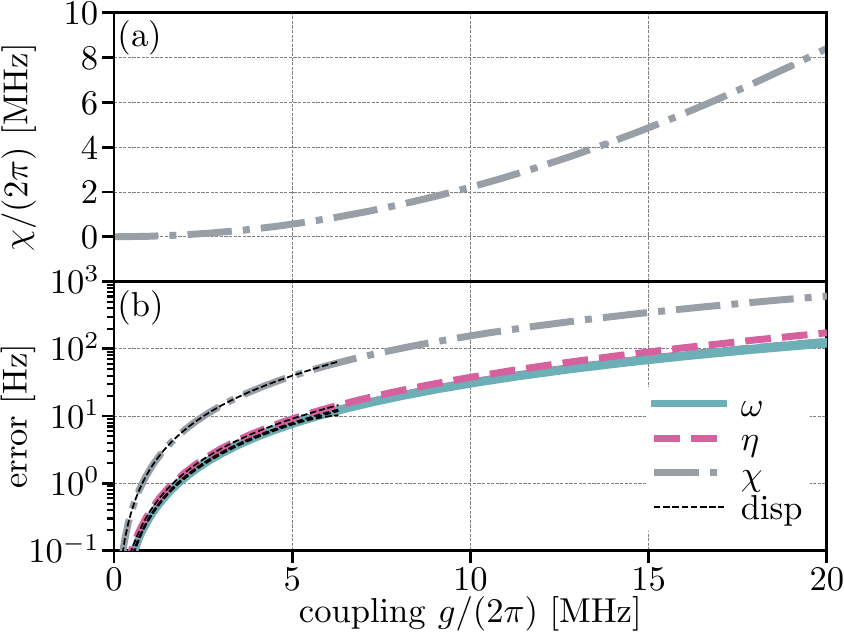}
\caption{Dispersive shift of capacitively-coupled transmons (a) and error on the frequencies, anharmonicities and frequency shift (b) at $25^\mathrm{th}$ order in $\xi$ for four transmon eigenstates as a function of coupling strength compared to numerical simulation with 30 Fock states for each transmon. 
The dashed lines correspond to the dispersive results, valid for small enough couplings.
The first  transmon is characterized by $E_C/h=200\,\mathrm{MHz}$ and $\xi_1=0.18$,  giving $\omega_1/(2\pi)=4234\,\mathrm{MHz}$, $\eta_1/(2\pi)=226\,\mathrm{MHz}$.
It is interacting with a coupling $g$ to
the second transmon    characterized by $E_C/h=200\,\mathrm{MHz}$ and $\xi_2=0.175$, giving $\omega_2/(2\pi)=4361\,\mathrm{MHz}$, $\eta_2/(2\pi)=225\,\mathrm{MHz}$.
}
\label{fig_coupled}
\end{figure}

For detunings much larger than the coupling, $|\omega_1-\omega_2|\gg g$, the transverse coupling gives rise to state-dependent frequency shifts.
At lowest order in the small parameter $g/|\omega_1-\omega_2|$, the frequencies and anharmonicities are modified as follows,
$\omega_{1,2}\to\omega_{1,2}+\delta\omega_{1,2}$ and $\eta_{1,2}\to\eta_{1,2}+\delta\eta_{1,2}$ with
\begin{align}
\delta\omega_{1,2}&=\pm\frac{g^2\lambda_1^2\lambda_2^2}{\omega_1-\omega_2},\\
\delta\eta_{1,2}&=2\delta\omega_{1,2}\mp\frac{2g^2\lambda_{2,1}^2\Lambda_{1,2}^2}{\omega_1-\omega_2\mp\eta_{1,2}}.
\end{align}
The interaction in the dispersive regime is of the form
$\hat{H}_\mathrm{disp}=\chi\ket{11}\bra{11}$ with the dispersive shift~\cite{Omalley15},
\begin{align}
\chi&=2g^2\left(\frac{\lambda_1^2\Lambda_2^2}{\omega_1-\omega_2+\eta_2}-\frac{\Lambda_1^2\lambda_2^2}{\omega_1-\omega_2-\eta_1}\right).
\end{align}
The dispersive parameters are compared to the numerical diagonalization of the coupled Hamiltonian Eq.~\eqref{H_coupled} in Fig.~\ref{fig_coupled}.

\section{Parametrically-activated entangling gates}
\label{secM}

\begin{figure}[t]
\includegraphics[width=\columnwidth]{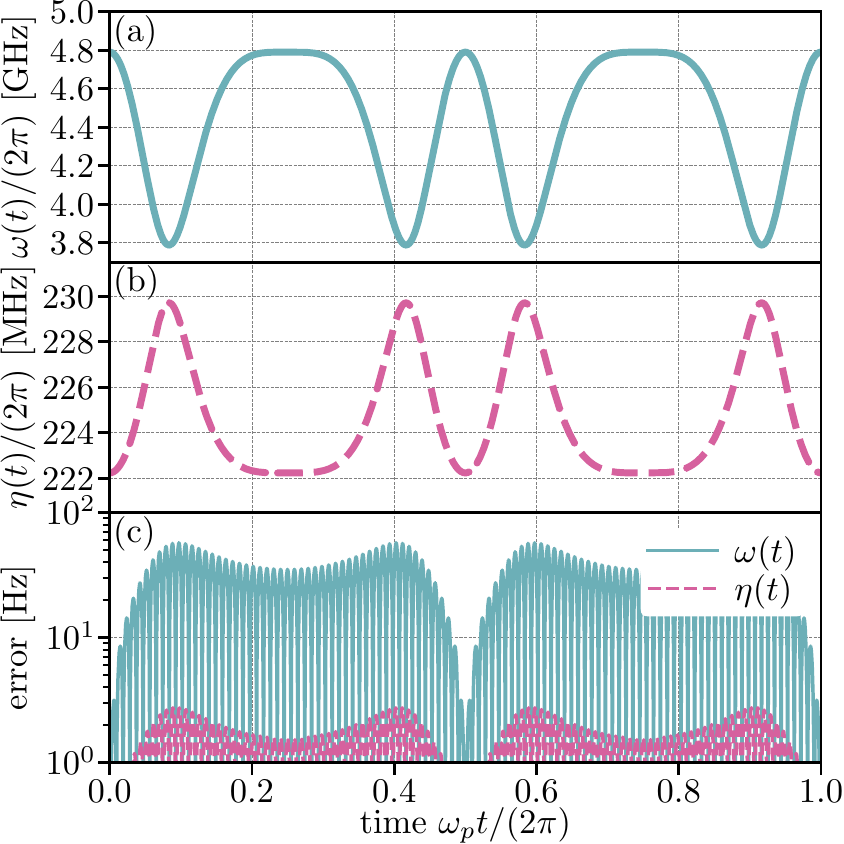}
\caption{Time evolution of the transmon frequency (a), anharmonicity (b) and error on the tunable transmon frequency and anharmonicity (c) under modulation at $25^\mathrm{th}$ order in $\xi$ as a function of time over a period of the modulation frequency, $2\pi/\omega_p$.
The time evolution of $\omega(t)$ and $\eta(t)$ from $\phi_\ext(t)=\phib_p+\phit_p\cos(\omega_pt+\theta_p)$ is compared to the Fourier series of the 50 first harmonics, Eq.~\eqref{pf}.
Parameters are the same as Fig.~\ref{fig_tunable} with $\phib_p=0$ (parking at a maximum) and $\phit_p=2\pi$ (modulating between three consecutive maxima).}
\label{fig_modulation}
\end{figure}

We consider two capacitively coupled transmon qubits, the first at fixed frequency and the second tunable.
The flux bias pulse is modulated; it renders the Hamiltonian time dependent via $E_{J_\eff}(t)$ and $\phi_\eff(t)$.
We note the parameters $\xi_F=\sqrt{2E_{C_F}/E_{J_F}}$ and $\xi_T(t)=\sqrt{2E_{C_T}/E_{J_\eff}(t)}$.
The coupling in the quadrature basis, $g(t)=g_C/(4\sqrt{\xi_F\xi_T(t)})$, is expressed in terms of the capacitive coupling $g_C$.
The system Hamiltonian in the transmon basis is then
\begin{multline}
\hat{H}_\mathrm{M}(t)
=\sum_{n=0}^2\left[E_{F_n}\hat{\Pi}_{F_n}+E_{T_n}(t)\hat{\Pi}_{T_n}\right]\\
+g(t)[\lambda_F\hat{\sigma}_{y_F}+\sqrt{2}\Lambda_F\hat{s}_{y_F}][\lambda_T(t)\hat{\sigma}_{y_T}+\sqrt{2}\Lambda_T(t)\hat{s}_{y_T}]\\
-\frac{\dot{\phi}_\eff(t)}{2\sqrt{\xi_T(t)}}\lambda_T(t)\hat{\sigma}_{y_T}
-\frac{\dot{\phi}_\eff(t)}{\sqrt{2\xi_T(t)}}\Lambda_T(t)\hat{s}_{y_T}
-\frac{\dot{\xi}_T(t)}{\xi_T(t)}\upsilon(t)\hat{S}_{y_T}.
\label{H2}
\end{multline}
We specify the modulation of the flux bias pulse,
\begin{align}
\phi_\ext(t)&=\phib_p+\phit_p\cos(\omega_pt+\theta_p),
\end{align}
which oscillates around the parking flux $\phib_p$ at the modulation frequency $\omega_p$ and amplitude $\phit_p$.
The Hamiltonian is time dependent via $\xi_T(t)$, therefore via $\cos[\phi_\ext(t)]$;
The general Fourier series of such time-dependent parameters, e.g.~the frequency $\omega_T(t)$, reads
\begin{align}
\omega_T(t)&=\sum_{k=0}^\infty\bm{\omega}_k\cos[k(\omega_pt+\theta_p)].
\label{pf}
\end{align}
The Fourier coefficients are defined as
$\bm{\omega}_k=\frac{\omega_p}{\pi(1+\delta_{k,0})}\int_0^{\frac{2\pi}{\omega_p}}\mathrm{d}t\,\cos[k(\omega_pt+\theta_p)]\omega_T(t)$.
An analytical expression of the Fourier components is provided in Appendix~\ref{App_FS}.
At the flux sweet spots, where $\cos\phib_p=0$ like at a maximum $\phib_p=0$ or a minimum $\phib_p=\pi$ of the energy spectrum, the odd Fourier coefficients vanish, $\bm{\omega}_{2k+1}=0$.
The leading behavior at small modulation amplitudes is then 
$\omega_T(t)\approx \bm{\omega}_0+\bm{\omega}_2\cos[2(\omega_pt+\theta_p)]$: 
a shifted frequency $\bm{\omega}_0$ corresponding to the averaged value of $\omega_T(t)$ and an oscillation of amplitude $\bm{\omega}_2$ at $2\omega_p$.
This upconversion from the flux modulation at~$\omega_p$ to the parameter modulation at~$2\omega_p$ comes from the shape of the spectrum around these parking points.
At an extremum, the slope vanishes and periodic excursions on the curvature doubles the frequency.
In the following, we consider parking at a flux sweet spot and use the notation 
$\overline{\omega}\equiv\bm{\omega}_0$, $\widetilde{\omega}\equiv\bm{\omega}_2$. 
The time evolution of the frequency is plotted in Fig.~\ref{fig_modulation}~(a) as a function of time over a period of modulation $2\pi/\omega_p$ for a large modulation amplitude, $\phit_p=2\pi$.
Starting from $\phi_\ext=0$, the flux goes through the minimum at $\phi_\ext=\pi$ and  bounces off at the maximum at $\phi_\ext=2\pi$, then comes back to the starting point before passing through the other minimum at $\phi_\ext=-\pi$ and reaching the maximum at $\phi_\ext=-2\pi$.
The temporal evolution of the anharmonicity is plotted in Fig.~\ref{fig_modulation}~(b).
The Fourier series with the 50 first harmonics is an accurate description of the frequency and anharmonicity for such large modulation amplitudes [see Fig.~\ref{fig_modulation}~(c)].

To highlight the parametrically-activated coupling, we go into the interaction picture with the unitary,
\begin{multline}
\hU_\mathrm{int}=\exp\bigg\{-i\int_0^t\mathrm{d}t'\,\sum_{n=0}^2[E_{F_n}\hat{\Pi}_{F_n}+E_{T_n}(t)\hat{\Pi}_{T_n}]\bigg\}.
\end{multline}
To proceed, we note the energy difference $\omega_{F,T_{ij}}=E_{F,T_j}-E_{F,T_i}$ 
and the time-integral $\varpi_{T_{ij}}(t)=\int_0^t\mathrm{d}t'\,\omega_{T_{ij}}(t')$.
The result Eq.~\eqref{FSexpf} is then used to find the Fourier series of the Hamiltonian in the interaction picture,
\begin{align}
\hat{H}_\mathrm{int}=&\sum_{n\in\mathbb{Z}}\bigg\{
g_{11}^{(n)}\Exp{i(2n\omega_p-\Delta)t}\Exp{i\beta_{01}^{(n)}}\,\ket{01}\bra{10}\nonumber\\
+&g_{21}^{(n)}\Exp{i(2n\omega_p-[\Delta-\eta_F])t}\Exp{i\beta_{01}^{(n)}}\,\ket{11}\bra{20}\nonumber\\
+&g_{12}^{(n)}\Exp{i(2n\omega_p-[\Delta+\eb_T])t}\Exp{i\beta_{12}^{(n)}}\,\ket{02}\bra{11}\nonumber\\
+&g_{22}^{(n)}\Exp{i(2n\omega_p-[\Delta-\eta_F+\eb_T])t}\Exp{i\beta_{12}^{(n)}}\,\ket{12}\bra{21}\nonumber\\
-&g_{11}^{(n)}\Exp{i(2n\omega_p+\Sigma)t}\Exp{i\beta_{01}^{(n)}}\,\ket{11}\bra{00}\nonumber\\
-&g_{21}^{(n)}\Exp{i(2n\omega_p+[\Sigma-\eta_F])t}\Exp{i\beta_{01}^{(n)}}\,\ket{21}\bra{10}\nonumber\\
-&g_{12}^{(n)}\Exp{i(2n\omega_p+[\Sigma-\eb_T])t}\Exp{i\beta_{12}^{(n)}}\,\ket{12}\bra{01}\nonumber\\
-&g_{22}^{(n)}\Exp{i(2n\omega_p+[\Sigma-\eta_F-\eb_T])t}\Exp{i\beta_{12}^{(n)}}\,\ket{22}\bra{11}\nonumber\\
+&\Omega_{01}^{(n)}\,\Exp{i((2n+1)\omega_p+\wb_{T_{01}})t}\Exp{i(\beta_{01}^{(n)}+\theta_p)}\ID\otimes\ket{1}\bra{0}\nonumber\\
+&\Omega_{12}^{(n)}\Exp{i((2n+1)\omega_p+[\wb_{T_{01}}-\eb_T])t}\Exp{i(\beta_{12}^{(n)}+\theta_p)}\ID\otimes\ket{2}\bra{1}\nonumber\\
+&\Omega_{02}^{(n)}\Exp{i(2n\omega_p+[2\wb_{T_{01}}-\eb_T])t}\Exp{i\beta_{02}^{(n)}}\ID\otimes\ket{2}\bra{0}
\bigg\}+\mathrm{h.c.}
\label{fullHint}
\end{align}
with the notation $\ket{FT}$
and where the frequency difference $\Delta$ and sum $\Sigma$ are,
\begin{align}
\Delta&=\omega_{F_{01}}-\wb_{T_{01}},&
\Sigma&=\omega_{F_{01}}+\wb_{T_{01}}.
\end{align}
Each line of the Hamiltonian Eq.~\eqref{fullHint} is a coupling between a pair of two-qubit states that can be brought into resonance with the right modulation frequency $\omega_p$.
The rate of the corresponding Rabi oscillations is set by the effective coupling strength, and hence the modulation amplitude. 
The capacitive coupling thus gives access to a large variety of two-qubit gates:
iSWAP with $\ket{10}\bra{01}$ at $\omega_p=\frac{1}{2}|\Delta|$;
controlled Z with $\ket{11}\bra{20}$ (CZ$_{20}$) at $\omega_p=\frac{1}{2}|\Delta-\eta_F|$ 
and $\ket{11}\bra{02}$ (CZ$_{02}$) at $\omega_p=\frac{1}{2}|\Delta+\eb_T|$;
Bell-Rabi with $\ket{00}\bra{11}$ at $\omega_p=\frac{1}{2}\Sigma$.
The flux modulation then completely controls the activation and rate of the gates.
The effective couplings read,
\begin{align}
g_{ij}^{(n)}&=\gb_{ij}\sum_{\substack{\{l_k\}\in\mathbb{Z}\\\sum_{k=1}^\infty kl_k=n}}\prod_{k=1}^\infty\BJ_{l_k}\!\left(\frac{\bm{[\omega_{j\textbf{-}1,j}]}_{2k}}{2k\omega_p}\right),
\label{effectivecoupling}
\end{align}
with the leading term equal to
$g_{ij}^{(n)}\simeq\gb_{ij}\BJ_n\!\left(\frac{\bm{[\omega_{j\textbf{-}1,j}]}_2}{2\omega_p}\right)$,
valid at small modulation amplitudes.
In Eq.~\eqref{effectivecoupling} we have taken into account only the averaged value of the couplings, $\bar{g}$, since the modulation amplitude of the couplings, $\tilde{g}$, is typically small.
The leading correction to the above expression for $g_{ij}^{(n)}$ reads 
$-\frac{1}{2}\gt_{ij}\{\BJ_{n-1}[\wt_{T_{j-1,j}}/(2\omega_p)]+\BJ_{n+1}[\wt_{T_{j-1,j}}/(2\omega_p)]\}$.
The time-independent parameters $\bar{g}$ and $\tilde{g}$ are obtained from the couplings
\begin{align}
g_{11}(t)&=g(t)\lambda_F\lambda_T(t),&
g_{21}(t)&=\sqrt{2}g(t)\Lambda_F\lambda_T(t),\\
g_{22}(t)&=2g(t)\Lambda_F\Lambda_T(t),&
g_{12}(t)&=\sqrt{2}g(t)\lambda_F\Lambda_T(t).
\end{align}
The phases are
$\beta_{ij_n}=2n\theta_p-\sum_{k=1}^\infty\frac{{\bm{[\omega_{ij}]}}_{2k}}{2k\omega_p}\sin(2k\theta_p)$.
The effective drives are given in Appendix~\ref{App_drives}.
For modulation frequencies well below qubit frequencies and close to half the frequency detuning~$\Delta$, a rotating wave approximation allows us to consider only the first three lines of Eq.~\eqref{fullHint}.

\begin{figure}[t]
\includegraphics[width=\columnwidth]{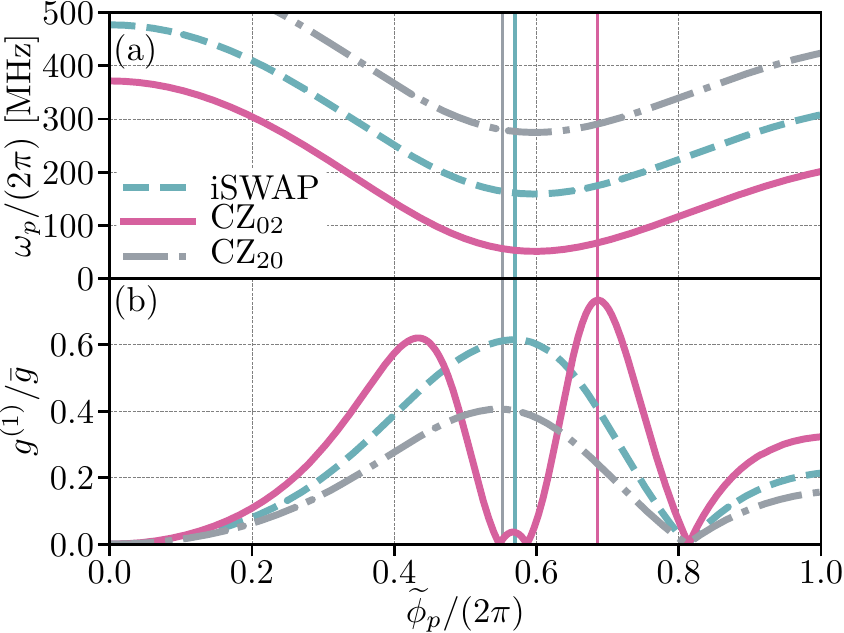}
\caption{
Modulation frequency activating three kinds of two-qubit gates (a) and corresponding effective coupling (b) as a function of the modulation amplitude.
A fixed transmon 
($E_C/h=200\,\mathrm{MHz}$, $\xi=0.21$,  giving $\omega/(2\pi)=3597\,\mathrm{MHz}$, $\eta/(2\pi)=232\,\mathrm{MHz}$)
is coupled to a tunable transmon 
($E_C/(2\pi)=190\,\mathrm{MHz}$, $\xi_\mathrm{max}=0.16$ and $\xi_\mathrm{min}=0.2$, giving $\omega_\mathrm{max}/(2\pi)=4551\,\mathrm{MHz}$, $\omega_\mathrm{min}/(2\pi)=3599\,\mathrm{MHz}$, $\eta_\mathrm{max}/(2\pi)=211\,\mathrm{MHz}$ and $\eta_\mathrm{min}/(2\pi)=218\,\mathrm{MHz}$) 
that is parked at a maximum of the energy spectrum ($\phib_p=0$).
The flux modulation amplitude $\phit_p$ and frequency $\omega_p$ are then chosen from (a) to activate a desired entangling gate among iSWAP, CZ$_{02}$ and CZ$_{20}$.
The bare coupling is renormalized by the modulation by a coefficient that is plotted in (b).
The vertical lines indicate operating points for fast gates.}
\label{fig_birch}
\end{figure}

The modulation frequencies that activate the entangling gates iSWAP, CZ$_{02}$, and CZ$_{20}$ are plotted in Fig.~\ref{fig_birch}~(a) as a function of the modulation amplitude.
The corresponding renormalization coefficient of the coupling is plotted in Fig.~\ref{fig_birch}~(b), it corresponds to $(g_{11}^{(1)}/\gb_{11},g_{12}^{(1)}/\gb_{12},g_{21}^{(1)}/\gb_{21})$ for (iSWAP,CZ$_{02}$,CZ$_{20}$), respectively.
Local maxima of the renormalization coefficient achieve $0.62,0.73,0.41$ in this example, they constitute the optimal points of operation for fast gates (vertical lines).

We have focused our study so far on the coherent dynamics of transmons under flux modulation.
Modulating the flux also changes the dissipative properties of tunable transmons, in particular dephasing due to flux noise~\cite{Didier}.
Indeed, even if the transmon is parked at a flux sweet spot where the qubit is flux-noise insensitive at leading order, when modulated the flux bias explores regions of higher dephasing rate thereby reducing the effective dephasing time~\cite{Rigetti-blue_2017}.
The physics of dephasing under flux modulation is the subject of ongoing research.

\section{Conclusion and perspectives}
\label{secD}

We have developed an accurate analytical model for coupled transmons in the presence of flux pulse modulation. Our approach provides efficient and precise simulations of large-scale transmon-based quantum processors, which would be otherwise intractable with numerical diagonalization. The tools developed in this work are useful for designing large-scale processors, optimizing design parameters and predicting the performances of quantum operations. This model has already been successfully used in recent experiments on parametrically-activated two-qubit gates~\cite{Rigetti-blue_2017,Rigetti-white_2017,Rigetti-clustering_2017}.

Moreover, our approach can be straightforwardly extended to the calculation of more than three transmon states to have more accurate expressions of frequency shifts. This is particularly important for finding the optimal regimes of operation of the controlled Z and iSWAP gates.

Developing an accurate model for dissipation in superconducting qubits is a natural extension of this work. The Keldysh formalism of Green's functions is a powerful framework particularly well suited for nonlinear open quantum systems~\cite{Kamenev_book,Lemonde_2013,Muller_2017}. 
The perturbative expansion of the transmon parameters can then be calculated in presence of a dissipative bath, allowing to take into account the relaxation and dephasing rates of the quantum machine in its optimization.

\section*{Acknowledgments}

We acknowledge fruitful daily collaborations with the experimental teams of Rigetti Computing.

\newpage
\appendix
\section{Diagonalization operator}
\label{App_Ueigen}

We provide the explicite expression of the diagonalization operator $\hU_\eigen$ at $5^\mathrm{th}$ order in $\xi$ of the first 5 transmon eigenstates from the first 25 Fock states.

\begin{align}
\hU_\eigen&=\sum_{p=0}^5\xi^p\hU_\eigen^{(p)},
\end{align}
%where $\hU_\eigen^{(0)}$ has 1 on the diagonal.
with (we write the conjugate operators for convenience),

\begin{align}
\hU_\eigen^{(0)\dag}
=
\begin{pmatrix}1 & 0 & 0 & 0 & 0\\ 0 & 1 & 0 & 0 & 0\\ 0 & 0 & 1 & 0 & 0\\ 0 & 0 & 0 & 1 & 0\\ 0 & 0 & 0 & 0 & 1\\ 0 & 0 & 0 & 0 & 0\\ 0 & 0 & 0 & 0 & 0\\ 0 & 0 & 0 & 0 & 0\\ 0 & 0 & 0 & 0 & 0\\ 0 & 0 & 0 & 0 & 0\\ 0 & 0 & 0 & 0 & 0\\ 0 & 0 & 0 & 0 & 0\\ 0 & 0 & 0 & 0 & 0\\ 0 & 0 & 0 & 0 & 0\\ 0 & 0 & 0 & 0 & 0\\ 0 & 0 & 0 & 0 & 0\\ 0 & 0 & 0 & 0 & 0\\ 0 & 0 & 0 & 0 & 0\\ 0 & 0 & 0 & 0 & 0\\ 0 & 0 & 0 & 0 & 0\\ 0 & 0 & 0 & 0 & 0\\ 0 & 0 & 0 & 0 & 0\\ 0 & 0 & 0 & 0 & 0\\ 0 & 0 & 0 & 0 & 0\\ 0 & 0 & 0 & 0 & 0\end{pmatrix},
\end{align}
\raggedbottom
\newpage

\begin{align}	
&\hU_\eigen^{(1)\dag}=
\frac{1}{2^{3}}\nonumber\\
&\times
\begin{pmatrix}0 & 0 & -\frac{1}{\sqrt{2}} & 0 & -\frac{1}{\sqrt{2}^{3}\sqrt{3}}\\ 0 & 0 & 0 & -\frac{5}{\sqrt{2}\sqrt{3}} & 0\\ \frac{1}{\sqrt{2}} & 0 & 0 & 0 & -\frac{7}{\sqrt{3}}\\ 0 & \frac{5}{\sqrt{2}\sqrt{3}} & 0 & 0 & 0\\ \frac{1}{\sqrt{2}^{3}\sqrt{3}} & 0 & \frac{7}{\sqrt{3}} & 0 & 0\\ 0 & \frac{\sqrt{5}}{\sqrt{2}^{3}\sqrt{3}} & 0 & \frac{3\sqrt{5}}{1} & 0\\ 0 & 0 & \frac{\sqrt{5}}{\sqrt{2}^{3}} & 0 & \frac{11\sqrt{5}}{\sqrt{2}\sqrt{3}}\\ 0 & 0 & 0 & \frac{\sqrt{35}}{\sqrt{2}^{3}\sqrt{3}} & 0\\ 0 & 0 & 0 & 0 & \frac{\sqrt{35}}{2\sqrt{3}}\\ 0 & 0 & 0 & 0 & 0\\ 0 & 0 & 0 & 0 & 0\\ 0 & 0 & 0 & 0 & 0\\ 0 & 0 & 0 & 0 & 0\\ 0 & 0 & 0 & 0 & 0\\ 0 & 0 & 0 & 0 & 0\\ 0 & 0 & 0 & 0 & 0\\ 0 & 0 & 0 & 0 & 0\\ 0 & 0 & 0 & 0 & 0\\ 0 & 0 & 0 & 0 & 0\\ 0 & 0 & 0 & 0 & 0\\ 0 & 0 & 0 & 0 & 0\\ 0 & 0 & 0 & 0 & 0\\ 0 & 0 & 0 & 0 & 0\\ 0 & 0 & 0 & 0 & 0\\ 0 & 0 & 0 & 0 & 0\end{pmatrix},
\end{align}
\raggedbottom

\begin{align}
&\hU_\eigen^{(2)\dag}=
\frac{1}{2^{2}}\nonumber\\
&\times
\begin{pmatrix}-\frac{13}{2^{8}3} & 0 & -\frac{5}{\sqrt{2}^{7}3} & 0 & \frac{\sqrt{3}}{\sqrt{2}^{9}}\\ 0 & -\frac{35}{2^{8}} & 0 & -\frac{13}{\sqrt{2}^{7}\sqrt{3}} & 0\\ \frac{13}{\sqrt{2}^{11}3} & 0 & -\frac{419}{2^{8}3} & 0 & -\frac{67\sqrt{3}}{2^{6}}\\ 0 & \frac{37}{\sqrt{2}^{11}\sqrt{3}} & 0 & -\frac{405}{2^{8}} & 0\\ \frac{1}{\sqrt{2}^{5}\sqrt{3}} & 0 & \frac{145}{2^{6}\sqrt{3}} & 0 & -\frac{961}{2^{8}}\\ 0 & \frac{41}{\sqrt{2}^{9}\sqrt{15}} & 0 & \frac{79\sqrt{5}}{2^{6}} & 0\\ \frac{23}{2^{6}3\sqrt{5}} & 0 & \frac{103}{\sqrt{2}^{7}3\sqrt{5}} & 0 & \frac{29\sqrt{15}}{\sqrt{2}^{9}}\\ 0 & \frac{11\sqrt{7}}{2^{6}\sqrt{5}} & 0 & \frac{103\sqrt{7}}{\sqrt{2}^{9}\sqrt{15}} & 0\\ \frac{\sqrt{35}}{\sqrt{2}^{15}3} & 0 & \frac{43\sqrt{7}}{2^{5}3\sqrt{5}} & 0 & \frac{3\sqrt{21}}{\sqrt{5}}\\ 0 & \frac{\sqrt{35}}{\sqrt{2}^{15}} & 0 & \frac{53\sqrt{7}}{2^{5}\sqrt{15}} & 0\\ 0 & 0 & \frac{5\sqrt{7}}{\sqrt{2}^{15}} & 0 & \frac{21\sqrt{21}}{\sqrt{2}^{11}}\\ 0 & 0 & 0 & \frac{5\sqrt{77}}{\sqrt{2}^{15}\sqrt{3}} & 0\\ 0 & 0 & 0 & 0 & \frac{5\sqrt{77}}{\sqrt{2}^{15}}\\ 0 & 0 & 0 & 0 & 0\\ 0 & 0 & 0 & 0 & 0\\ 0 & 0 & 0 & 0 & 0\\ 0 & 0 & 0 & 0 & 0\\ 0 & 0 & 0 & 0 & 0\\ 0 & 0 & 0 & 0 & 0\\ 0 & 0 & 0 & 0 & 0\\ 0 & 0 & 0 & 0 & 0\\ 0 & 0 & 0 & 0 & 0\\ 0 & 0 & 0 & 0 & 0\\ 0 & 0 & 0 & 0 & 0\\ 0 & 0 & 0 & 0 & 0\end{pmatrix},
\end{align}
\raggedbottom

\begin{align}
&\hU_\eigen^{(3)\dag}=
\frac{1}{2^{8}}\nonumber\\
&\times
\begin{pmatrix}-\frac{17}{2^{3}3} & 0 & -\frac{725}{\sqrt{2}^{11}3} & 0 & \frac{1747}{\sqrt{2}^{13}\sqrt{3}}\\ 0 & -\frac{113}{2^{2}3} & 0 & -\frac{869\sqrt{3}}{\sqrt{2}^{11}} & 0\\ \frac{347}{\sqrt{2}^{11}3} & 0 & -\frac{1241}{2^{3}3} & 0 & -\frac{6217}{2^{5}\sqrt{3}}\\ 0 & \frac{515\sqrt{3}}{\sqrt{2}^{11}} & 0 & -\frac{189}{1} & 0\\ \frac{605}{\sqrt{2}^{13}\sqrt{3}} & 0 & \frac{3793}{2^{5}\sqrt{3}} & 0 & -\frac{2165}{2^{2}}\\ 0 & \frac{2009\sqrt{5}}{\sqrt{2}^{13}\sqrt{3}} & 0 & \frac{2225\sqrt{5}}{2^{5}} & 0\\ \frac{409}{2^{4}3\sqrt{5}} & 0 & \frac{68981}{\sqrt{2}^{13}3\sqrt{5}} & 0 & \frac{44689}{\sqrt{2}^{11}\sqrt{15}}\\ 0 & \frac{461\sqrt{5}}{2^{4}\sqrt{7}} & 0 & \frac{8663\sqrt{35}}{\sqrt{2}^{13}\sqrt{3}} & 0\\ \frac{239}{\sqrt{2}^{5}3\sqrt{35}} & 0 & \frac{52681}{2^{5}3\sqrt{35}} & 0 & \frac{36119\sqrt{7}}{2^{5}\sqrt{15}}\\ 0 & \frac{127\sqrt{5}}{\sqrt{2}^{3}3\sqrt{7}} & 0 & \frac{52763\sqrt{5}}{2^{5}3\sqrt{21}} & 0\\ \frac{31\sqrt{7}}{2^{5}3} & 0 & \frac{1963}{\sqrt{2}^{5}3\sqrt{7}} & 0 & \frac{202283}{\sqrt{2}^{9}3\sqrt{21}}\\ 0 & \frac{41\sqrt{77}}{2^{5}3} & 0 & \frac{233\sqrt{11}}{\sqrt{2}\sqrt{21}} & 0\\ \frac{5\sqrt{77}}{2^{6}3\sqrt{3}} & 0 & \frac{17\sqrt{231}}{\sqrt{2}^{9}} & 0 & \frac{3769\sqrt{11}}{\sqrt{2}^{5}3\sqrt{7}}\\ 0 & \frac{5\sqrt{1001}}{2^{6}3\sqrt{3}} & 0 & \frac{61\sqrt{1001}}{\sqrt{2}^{9}3} & 0\\ 0 & 0 & \frac{35\sqrt{143}}{2^{6}3\sqrt{3}} & 0 & \frac{497\sqrt{143}}{2^{5}3}\\ 0 & 0 & 0 & \frac{35\sqrt{715}}{2^{6}3\sqrt{3}} & 0\\ 0 & 0 & 0 & 0 & \frac{35\sqrt{715}}{2^{5}3\sqrt{3}}\\ 0 & 0 & 0 & 0 & 0\\ 0 & 0 & 0 & 0 & 0\\ 0 & 0 & 0 & 0 & 0\\ 0 & 0 & 0 & 0 & 0\\ 0 & 0 & 0 & 0 & 0\\ 0 & 0 & 0 & 0 & 0\\ 0 & 0 & 0 & 0 & 0\\ 0 & 0 & 0 & 0 & 0\end{pmatrix},
\end{align}
\raggedbottom

\begin{widetext}
\begin{align}
&\hU_\eigen^{(4)\dag}=
\frac{1}{\sqrt{2}^{21}}\nonumber\\
&\times
\begin{pmatrix}-\frac{62333}{\sqrt{2}^{21}15} & 0 & -\frac{9911}{2^{5}15} & 0 & \frac{39007}{2^{6}3\sqrt{3}}\\ 0 & -\frac{1313189}{\sqrt{2}^{21}15} & 0 & -\frac{137483}{2^{5}15\sqrt{3}} & 0\\ \frac{16181}{2^{7}15} & 0 & -\frac{651579}{\sqrt{2}^{21}} & 0 & -\frac{3086339}{\sqrt{2}^{15}15\sqrt{3}}\\ 0 & \frac{110401}{2^{7}5\sqrt{3}} & 0 & -\frac{44292461}{\sqrt{2}^{21}15} & 0\\ \frac{269\sqrt{3}}{2^{5}} & 0 & \frac{227829\sqrt{3}}{\sqrt{2}^{15}5} & 0 & -\frac{149454457}{\sqrt{2}^{21}15}\\ 0 & \frac{8281\sqrt{5}}{2^{6}\sqrt{3}} & 0 & \frac{254291\sqrt{5}}{\sqrt{2}^{15}3} & 0\\ \frac{7601}{\sqrt{2}^{15}\sqrt{5}} & 0 & \frac{5833}{3\sqrt{5}} & 0 & \frac{55231}{2^{2}3\sqrt{15}}\\ 0 & \frac{85841\sqrt{7}}{\sqrt{2}^{15}3\sqrt{5}} & 0 & \frac{5911741}{2^{6}3\sqrt{105}} & 0\\ \frac{164443}{2^{9}3\sqrt{35}} & 0 & \frac{1543667}{\sqrt{2}^{13}3\sqrt{35}} & 0 & \frac{11114269}{\sqrt{2}^{11}3\sqrt{105}}\\ 0 & \frac{4110871}{2^{9}9\sqrt{35}} & 0 & \frac{1935007\sqrt{5}}{\sqrt{2}^{13}3\sqrt{21}} & 0\\ \frac{135697}{\sqrt{2}^{13}45\sqrt{7}} & 0 & \frac{6278647\sqrt{7}}{2^{9}45} & 0 & \frac{1265399\sqrt{7}}{2^{6}3\sqrt{3}}\\ 0 & \frac{283201\sqrt{11}}{\sqrt{2}^{13}45\sqrt{7}} & 0 & \frac{26982119\sqrt{11}}{2^{9}15\sqrt{21}} & 0\\ \frac{2759\sqrt{11}}{\sqrt{2}^{9}15\sqrt{21}} & 0 & \frac{199855\sqrt{11}}{2^{7}3\sqrt{21}} & 0 & \frac{135190589\sqrt{11}}{2^{9}45\sqrt{7}}\\ 0 & \frac{8843\sqrt{143}}{\sqrt{2}^{11}15\sqrt{21}} & 0 & \frac{1592063\sqrt{143}}{2^{7}45\sqrt{7}} & 0\\ \frac{91\sqrt{143}}{2^{7}3\sqrt{3}} & 0 & \frac{3217\sqrt{143}}{\sqrt{2}^{7}15\sqrt{3}} & 0 & \frac{73963\sqrt{143}}{\sqrt{2}^{5}45}\\ 0 & \frac{343\sqrt{715}}{2^{7}9} & 0 & \frac{17593\sqrt{143}}{\sqrt{2}^{11}3\sqrt{15}} & 0\\ \frac{35\sqrt{715}}{2^{9}9} & 0 & \frac{413\sqrt{715}}{\sqrt{2}^{11}9} & 0 & \frac{11509\sqrt{143}}{\sqrt{2}^{7}3\sqrt{15}}\\ 0 & \frac{35\sqrt{12155}}{2^{9}9} & 0 & \frac{161\sqrt{12155}}{\sqrt{2}^{11}3\sqrt{3}} & 0\\ 0 & 0 & \frac{35\sqrt{12155}}{2^{9}3} & 0 & \frac{553\sqrt{12155}}{2^{6}3\sqrt{3}}\\ 0 & 0 & 0 & \frac{35\sqrt{230945}}{2^{9}3\sqrt{3}} & 0\\ 0 & 0 & 0 & 0 & \frac{175\sqrt{46189}}{2^{9}3\sqrt{3}}\\ 0 & 0 & 0 & 0 & 0\\ 0 & 0 & 0 & 0 & 0\\ 0 & 0 & 0 & 0 & 0\\ 0 & 0 & 0 & 0 & 0\end{pmatrix},
\end{align}
\end{widetext}
\raggedbottom

\begin{widetext}
\begin{align}
&\hU_\eigen^{(5)\dag}=
\frac{1}{\sqrt{2}^{33}}\nonumber\\
&\times
\begin{pmatrix}-\frac{153917}{\sqrt{2}^{9}45} & 0 & -\frac{10765201}{2^{8}45} & 0 & \frac{202767029}{2^{9}15\sqrt{3}}\\ 0 & -\frac{41527}{\sqrt{2}^{3}3} & 0 & -\frac{10916291}{2^{8}3\sqrt{3}} & 0\\ \frac{4600889}{2^{8}45} & 0 & -\frac{50029111}{\sqrt{2}^{9}45} & 0 & -\frac{51091643}{\sqrt{2}^{15}5\sqrt{3}}\\ 0 & \frac{9434411}{2^{8}3\sqrt{3}} & 0 & -\frac{28185541}{\sqrt{2}^{7}9} & 0\\ \frac{9851179}{2^{9}15\sqrt{3}} & 0 & \frac{188993981}{\sqrt{2}^{15}15\sqrt{3}} & 0 & -\frac{187170527}{\sqrt{2}^{7}15}\\ 0 & \frac{25184897}{2^{9}\sqrt{15}} & 0 & \frac{394451371}{\sqrt{2}^{15}9\sqrt{5}} & 0\\ \frac{994363}{\sqrt{2}^{11}9\sqrt{5}} & 0 & \frac{887375369}{2^{9}9\sqrt{5}} & 0 & \frac{204498013}{2^{8}3\sqrt{15}}\\ 0 & \frac{1873273\sqrt{7}}{\sqrt{2}^{11}3\sqrt{5}} & 0 & \frac{5612971213}{2^{9}3\sqrt{105}} & 0\\ \frac{558181}{2^{3}9\sqrt{35}} & 0 & \frac{530587109}{\sqrt{2}^{13}9\sqrt{35}} & 0 & \frac{3942561341}{\sqrt{2}^{17}\sqrt{105}}\\ 0 & \frac{350365\sqrt{5}}{29\sqrt{7}} & 0 & \frac{451113919}{\sqrt{2}^{13}\sqrt{105}} & 0\\ \frac{4552733}{\sqrt{2}^{13}15\sqrt{7}} & 0 & \frac{99431659}{2^{3}45\sqrt{7}} & 0 & \frac{7205463341}{2^{6}15\sqrt{21}}\\ 0 & \frac{94247731}{\sqrt{2}^{13}9\sqrt{77}} & 0 & \frac{7581187\sqrt{11}}{2^{2}3\sqrt{21}} & 0\\ \frac{117428699}{\sqrt{2}^{15}45\sqrt{231}} & 0 & \frac{3349510421}{2^{6}45\sqrt{231}} & 0 & \frac{301458499\sqrt{11}}{2^{4}15\sqrt{7}}\\ 0 & \frac{6249499\sqrt{13}}{\sqrt{2}^{15}\sqrt{231}} & 0 & \frac{449418449\sqrt{13}}{2^{6}9\sqrt{77}} & 0\\ \frac{108659\sqrt{143}}{2^{5}45\sqrt{3}} & 0 & \frac{555486781\sqrt{13}}{\sqrt{2}^{15}45\sqrt{33}} & 0 & \frac{1352574373\sqrt{13}}{\sqrt{2}^{13}15\sqrt{11}}\\ 0 & \frac{203167\sqrt{143}}{2^{5}9\sqrt{5}} & 0 & \frac{108310393\sqrt{13}}{\sqrt{2}^{15}\sqrt{165}} & 0\\ \frac{2633\sqrt{143}}{2^{3}9\sqrt{5}} & 0 & \frac{59661\sqrt{715}}{\sqrt{2}^{13}} & 0 & \frac{1051619747\sqrt{13}}{\sqrt{2}^{15}3\sqrt{165}}\\ 0 & \frac{491\sqrt{2431}}{9\sqrt{5}} & 0 & \frac{454661\sqrt{2431}}{\sqrt{2}^{13}\sqrt{15}} & 0\\ \frac{329\sqrt{12155}}{\sqrt{2}^{13}9} & 0 & \frac{16369\sqrt{2431}}{2^{3}9\sqrt{5}} & 0 & \frac{2945699\sqrt{2431}}{2^{6}3\sqrt{15}}\\ 0 & \frac{133\sqrt{230945}}{\sqrt{2}^{13}3} & 0 & \frac{401\sqrt{138567}}{2^{2}\sqrt{5}} & 0\\ \frac{35\sqrt{46189}}{\sqrt{2}^{15}9} & 0 & \frac{2345\sqrt{46189}}{2^{6}9} & 0 & \frac{3071\sqrt{46189}}{2^{3}\sqrt{3}}\\ 0 & \frac{35\sqrt{323323}}{\sqrt{2}^{15}3\sqrt{3}} & 0 & \frac{2695\sqrt{323323}}{2^{6}9} & 0\\ 0 & 0 & \frac{385\sqrt{29393}}{\sqrt{2}^{15}3\sqrt{3}} & 0 & \frac{11165\sqrt{29393}}{\sqrt{2}^{13}3}\\ 0 & 0 & 0 & \frac{385\sqrt{676039}}{\sqrt{2}^{15}9} & 0\\ 0 & 0 & 0 & 0 & \frac{385\sqrt{676039}}{2^{7}3\sqrt{3}}\end{pmatrix}.
\end{align}
\end{widetext}
\raggedbottom

\newpage\mbox{}\newpage\mbox{}\newpage
\section{Transmon parameters at $25^\mathrm{th}$ order in $\xi$}
\label{App_25}

Transmon frequency.

\begin{align}
\omega&=\sqrt{8E_CE_J}\nonumber\\
&-E_C\bigg[
1
+\frac{1}{2^{2}}\xi
+\frac{21}{2^{7}}\xi^{2}
+\frac{19}{2^{7}}\xi^{3}
+\frac{5319}{2^{15}}\xi^{4}\nonumber\\
&+\frac{6649}{2^{15}}\xi^{5}\nonumber\\
&+\frac{1180581}{2^{22}}\xi^{6}\nonumber\\
&+\frac{446287}{2^{20}}\xi^{7}\nonumber\\
&+\frac{1489138635}{2^{31}}\xi^{8}\nonumber\\
&+\frac{648381403}{2^{29}}\xi^{9}\nonumber\\
&+\frac{614557854099}{2^{38}}\xi^{10}\nonumber\\
&+\frac{75265839129}{2^{34}}\xi^{11}\nonumber\\
&+\frac{637411859250147}{2^{46}}\xi^{12}\nonumber\\
&+\frac{86690561488017}{2^{42}}\xi^{13}\nonumber\\
&+\frac{405768570324517701}{2^{53}}\xi^{14}\nonumber\\
&+\frac{15191635582891041}{2^{47}}\xi^{15}\nonumber\\
&+\frac{2497063196283456607731}{2^{63}}\xi^{16}\nonumber\\
&+\frac{102281923716042917215}{2^{57}}\xi^{17}\nonumber\\
&+\frac{2292687293949773041433127}{2^{70}}\xi^{18}\nonumber\\
&+\frac{25544408245062216574759}{2^{62}}\xi^{19}\nonumber\\
&+\frac{4971071120163260007203175705}{2^{78}}\xi^{20}\nonumber\\
&+\frac{59956026877695226936825271}{2^{70}}\xi^{21}\nonumber\\
&+\frac{6299936888270974385982624367587}{2^{85}}\xi^{22}\nonumber\\
&+\frac{20465345194746565030172477629}{2^{75}}\xi^{23}\nonumber\\
&+\frac{36984324599399309412347250837528543}{2^{94}}\xi^{24}
\bigg].
\end{align}

\newpage
Transmon anharmonicity.
\begin{align}
\eta&=E_C
\bigg[
1
+\frac{9}{2^{4}}\xi
+\frac{81}{2^{7}}\xi^{2}
+\frac{3645}{2^{12}}\xi^{3}
+\frac{46899}{2^{15}}\xi^{4}\nonumber\\
&+\frac{1329129}{2^{19}}\xi^{5}\nonumber\\
&+\frac{20321361}{2^{22}}\xi^{6}\nonumber\\
&+\frac{2648273373}{2^{28}}\xi^{7}\nonumber\\
&+\frac{45579861135}{2^{31}}\xi^{8}\nonumber\\
&+\frac{1647988255539}{2^{35}}\xi^{9}\nonumber\\
&+\frac{31160327412879}{2^{38}}\xi^{10}\nonumber\\
&+\frac{2457206583272505}{2^{43}}\xi^{11}\nonumber\\
&+\frac{50387904068904927}{2^{46}}\xi^{12}\nonumber\\
&+\frac{2145673984043982897}{2^{50}}\xi^{13}\nonumber\\
&+\frac{47368663010124907041}{2^{53}}\xi^{14}\nonumber\\
&+\frac{17329540083222030375645}{2^{60}}\xi^{15}\nonumber\\
&+\frac{410048712835835979799431}{2^{63}}\xi^{16}\nonumber\\
&+\frac{20066784213453521778111375}{2^{67}}\xi^{17}\nonumber\\
&+\frac{507447585299180759749453827}{2^{70}}\xi^{18}\nonumber\\
&+\frac{53019019946496461235728807475}{2^{75}}\xi^{19}\nonumber\\
&+\frac{1429754157181172012054040903645}{2^{78}}\xi^{20}\nonumber\\
&+\frac{79571741391885949104006842758911}{2^{82}}\xi^{21}\nonumber\\
&+\frac{2283773190022904454409743892590327}{2^{85}}\xi^{22}\nonumber\\
&+\frac{540565733415401595950277192471356985}{2^{91}}\xi^{23}\nonumber\\
&+\frac{16479511149218202447739080120870460083}{2^{94}}\xi^{24}
\bigg].
\end{align}

\newpage
Weight of the charge operator $\hat{N}$ on $\hs_y$.
\begin{align}
\lambda&=1-\bigg[
\frac{1}{2^{3}}\xi
+\frac{11}{2^{8}}\xi^{2}
+\frac{65}{2^{11}}\xi^{3}
+\frac{4203}{2^{17}}\xi^{4}\nonumber\\
&+\frac{40721}{2^{20}}\xi^{5}\nonumber\\
&+\frac{1784885}{2^{25}}\xi^{6}\nonumber\\
&+\frac{21465147}{2^{28}}\xi^{7}\nonumber\\
&+\frac{4455462653}{2^{35}}\xi^{8}\nonumber\\
&+\frac{61698199851}{2^{38}}\xi^{9}\nonumber\\
&+\frac{3623317643901}{2^{43}}\xi^{10}\nonumber\\
&+\frac{56143119646191}{2^{46}}\xi^{11}\nonumber\\
&+\frac{7321743985484303}{2^{52}}\xi^{12}\nonumber\\
&+\frac{125280019793719221}{2^{55}}\xi^{13}\nonumber\\
&+\frac{8984438512815167237}{2^{60}}\xi^{14}\nonumber\\
&+\frac{168544684286400995331}{2^{63}}\xi^{15}\nonumber\\
&+\frac{105741913308715347076701}{2^{71}}\xi^{16}\nonumber\\
&+\frac{2164311753394257835891059}{2^{74}}\xi^{17}\nonumber\\
&+\frac{184798694135089048676718297}{2^{79}}\xi^{18}\nonumber\\
&+\frac{4109869091672376619457585371}{2^{82}}\xi^{19}\nonumber\\
&+\frac{761062061371895548979377743237}{2^{88}}\xi^{20}\nonumber\\
&+\frac{18317012159331390907042783219855}{2^{91}}\xi^{21}\nonumber\\
&+\frac{1831630981593132690479908285273395}{2^{96}}\xi^{22}\nonumber\\
&+\frac{47512263370928552970648689915451821}{2^{99}}\xi^{23}\nonumber\\
&+\frac{20440707519371829420653298425077482201}{2^{106}}\xi^{24}\nonumber\\
&+\frac{569157711742925565406447462105395143103}{2^{109}}\xi^{25}
\bigg].
\end{align}

\newpage
Weight of the charge operator $\hat{N}$ on $\hS_y$.
\begin{align}
\Lambda&=1-\bigg[
\frac{1}{2^{2}}\xi
+\frac{73}{2^{9}}\xi^{2}
+\frac{79}{2^{9}}\xi^{3}
+\frac{113685}{2^{19}}\xi^{4}\nonumber \\
&+\frac{747533}{2^{21}}\xi^{5}\nonumber \\
&+\frac{175422349}{2^{28}}\xi^{6}\nonumber \\
&+\frac{698471247}{2^{29}}\xi^{7}\nonumber \\
&+\frac{1520876829389}{2^{39}}\xi^{8}\nonumber \\
&+\frac{13668058962903}{2^{41}}\xi^{9}\nonumber \\
&+\frac{4122722770459287}{2^{48}}\xi^{10}\nonumber \\
&+\frac{2534488707574995}{2^{46}}\xi^{11}\nonumber \\
&+\frac{26543348405245135937}{2^{58}}\xi^{12}\nonumber \\
&+\frac{281548290669062665101}{2^{60}}\xi^{13}\nonumber \\
&+\frac{98933257452818263360213}{2^{67}}\xi^{14}\nonumber \\
&+\frac{561603848629069641896937}{2^{68}}\xi^{15}\nonumber \\
&+\frac{3372037991404912212166296765}{2^{79}}\xi^{16}\nonumber \\
&+\frac{40819563311626093062783992331}{2^{81}}\xi^{17}\nonumber \\
&+\frac{16314102788878455728540034311379}{2^{88}}\xi^{18}\nonumber \\
&+\frac{52535388424912627194648863334467}{2^{88}}\xi^{19}\nonumber \\
&+\frac{178610931461508948221684711385383067}{2^{98}}\xi^{20}\nonumber \\
&+\frac{2444937960639526361173164055382471707}{2^{100}}\xi^{21}\nonumber \\
&+\frac{1103567409503040799217165335410059740779}{2^{107}}\xi^{22}\nonumber \\
&+\frac{8017554417550804194373089101907638666069}{2^{108}}\xi^{23}\nonumber \\
&+\frac{30711842188423912661533983529887505235301321}{2^{118}}\xi^{24}\nonumber \\
&+\frac{473069922042437374183190305740304564254754227}{2^{120}}\xi^{25}
\bigg].
\end{align}

\newpage
Weight of the nonadiabatic term.
\begin{align}
\upsilon&=-\sqrt{2}\bigg[
\frac{1}{2^{4}}\xi
+\frac{11}{2^{8}}\xi^{2}
+\frac{321}{2^{13}}\xi^{3}
+\frac{5609}{2^{17}}\xi^{4}\nonumber\\
&+\frac{450555}{2^{23}}\xi^{5}\nonumber\\
&+\frac{10164565}{2^{27}}\xi^{6}\nonumber\\
&+\frac{507453429}{2^{32}}\xi^{7}\nonumber\\
&+\frac{13856203441}{2^{36}}\xi^{8}\nonumber\\
&+\frac{3280643089875}{2^{43}}\xi^{9}\nonumber\\
&+\frac{104433423564937}{2^{47}}\xi^{10}\nonumber\\
&+\frac{7105628334651135}{2^{52}}\xi^{11}\nonumber\\
&+\frac{256923396012609391}{2^{56}}\xi^{12}\nonumber\\
&+\frac{39309225873672019119}{2^{62}}\xi^{13}\nonumber\\
&+\frac{1584176336386469903609}{2^{66}}\xi^{14}\nonumber\\
&+\frac{134062942734813033556893}{2^{71}}\xi^{15}\nonumber\\
&+\frac{5937992825016447235650113}{2^{75}}\xi^{16}\nonumber\\
&+\frac{4393462009358111483920628355}{2^{83}}\xi^{17}\nonumber\\
&+\frac{211630177923548593260384339985}{2^{87}}\xi^{18}\nonumber\\
&+\frac{21195084297362748051328855644603}{2^{92}}\xi^{19}\nonumber\\
&+\frac{1101441422698682678884159890620131}{2^{96}}\xi^{20}\nonumber\\
&+\frac{237236307127374537401655462955710741}{2^{102}}\xi^{21}\nonumber\\
&+\frac{13218681516317907311568006522672236075}{2^{106}}\xi^{22}\nonumber\\
&+\frac{1522482900088767896105176250210633085315}{2^{111}}\xi^{23}\nonumber\\
&+\frac{90520992079359034852853176891693012642775}{2^{115}}\xi^{24}\nonumber\\
&+\frac{44409876028541673056803493111783651485068951}{2^{122}}\xi^{25}
\bigg].
\end{align}

\newpage
\section{Fourier series}
\label{App_FS}

A parameter $f(t)$ that depends on time via $\xi_T(t)$ can be written as $f(t)=F[\Xi(\cos[\phi_\ext(t)])]$
where $\Xi=\xi_T^{-4}$ is linear in $\cos[\phi_\ext(t)]$.
The external flux is modulated at the frequency $\omega_p$, $\phi_\ext(t)=\phib_p+\phit_p\cos(\omega_pt+\theta_p)$.
We use the identity $\Exp{iy\sin x}=\sum_{n\in\mathbb{Z}}\BJ_n(y)\Exp{inx}$ to calculate the general Fourier series of $f$ in harmonics $\bm{f}_k$ of the modulation frequency $\omega_p$,
\begin{align}
f(t)&=\sum_{k=0}^\infty\bm{f}_k\cos[k(\omega_pt+\theta_p)],\label{FSf}\\
\bm{f}_k&=\sum_{n=0}^\infty\frac{\widetilde{\Xi}^n}{n!}F^{(n)}(\overline{\Xi})S_{k,n},\label{series}\\
S_{k,n}&=(2-\delta_{k,0})\frac{1}{2^n}\sum_{j=0}^{\mathrm{fl}(\frac{n}{2})}(2-\delta_{2j,n})\,s_{k,n,j},\\
s_{k,n,j}&=\binom{n}{j}\cos[(n-2j)\phib_p+k\tfrac{\pi}{2}]\,\BJ_k[(n-2j)\phit_p],
\end{align}
with $\delta_{x,y}$ the Kronecker delta function, $\mathrm{fl}$ the floor function,
$\overline{\Xi}=(E_{J_1}^2+E_{J_2}^2)/(4E_C^2)$ and $\widetilde{\Xi}=E_{J_1}E_{J_2}/(2E_C^2)$.
At flux sweet spots, $\phib_p=m\pi$, $m\in\mathbb{Z}$, the odd Fourier coefficients vanish, $\mathbf{f}_{2k+1}=0$.

To further simplify the Fourier coefficients,
we use the expression of $f$ obtained from perturbation theory $f=\sum_{p\in\mathbb{Z}}f^{(p)}\xi^p$.
The sum on $p$ starts at $p=-1$ for $\omega$ because $\omega_h=4E_C/\xi$, at $p=0$ for $\eta$.
The substitution yields,
\begin{align}
\bm{f}_k&=\sum_{n=0}^\infty\frac{(-\frac{1}{2}\mathcal{X})^n}{n!}
\cos(n\phib_p+k\tfrac{\pi}{2})\BJ_k(n\phit_p)S_{k,n},\\
S_{k,n}&=(2-\delta_{k,0})(2-\delta_{n,0})\sum_{p\in\mathbb{Z}}s_{k,n,p},\\
s_{k,n,p}&=f^{(p)}\bar{\xi}^pR_{n,p}\,{}_2\mathrm{F}_1(\tfrac{n}{2}\!+\!\tfrac{p}{8},\tfrac{n+1}{2}\!+\!\tfrac{p}{8},n\!+\!1,\mathcal{X}^2),
\end{align}
with $\mathcal{X}=\widetilde{\Xi}/\overline{\Xi}$, $\bar{\xi}=\overline{\Xi}^{-\frac{1}{4}}$, ${}_2\mathrm{F}_1$ the hypergeometric function,
$R_{n,p}=\left\{\begin{array}{ll}
0&\mathrm{if}\ p=0\ \mathrm{and}\ n>0\\
\frac{\Gamma(n+\frac{p}{4})}{\Gamma(\frac{p}{4})}&\mathrm{else}
\end{array}\right.$ and $\Gamma$ the gamma function.

We then calculate the Fourier series of $\Exp{i\int_0^t\mathrm{d}t'f(t')}$,
\begin{align}
\Exp{i\int_0^t\mathrm{d}t'f(t')}&=\Exp{i(\bm{f}_0t-\Theta_p)}\sum_{n\in\mathbb{Z}}\bm{\varepsilon}_n\Exp{in(\omega_pt+\theta_p)},\label{FSexpf}\\
\bm{\varepsilon}_n&=\sum_{\substack{\{l_k\}\in\mathbb{Z}\\\sum_{k=1}^\infty kl_k=n}}\prod_{k=1}^\infty\BJ_{l_k}\!\left(\frac{\bm{f}_k}{k\omega_p}\right),
\end{align}
with the phase
$\displaystyle \Theta_p=\sum_{k=1}^\infty\frac{\bm{f}_k}{k\omega_p}\sin(k\theta_p)$.

\section{Effective drives}
\label{App_drives}

The expression of the effective drives in Eq.~\eqref{fullHint} is
\begin{align}
\Omega_{01}^{(n)}&=(\nb_{01}-\tfrac{1}{2}\nt_{01})\left[\BJ_{n}\!\left(\frac{\wt_{T_{01}}}{2\omega_p}\right)+\BJ_{n+1}\!\left(\frac{\wt_{T_{01}}}{2\omega_p}\right)\right]\nonumber\\
&-\tfrac{1}{2}\nt_{01}\left[\BJ_{n-1}\!\left(\frac{\wt_{T_{01}}}{2\omega_p}\right)+\BJ_{n+2}\!\left(\frac{\wt_{T_{01}}}{2\omega_p}\right)\right],\\
\Omega_{12}^{(n)}&=(\nb_{12}-\tfrac{1}{2}\nt_{12})\left[\BJ_{n}\!\left(\frac{\wt_{T_{12}}}{2\omega_p}\right)+\BJ_{n+1}\!\left(\frac{\wt_{T_{12}}}{2\omega_p}\right)\right]\nonumber\\
&-\tfrac{1}{2}\nt_{12}\left[\BJ_{n-1}\!\left(\frac{\wt_{T_{12}}}{2\omega_p}\right)+\BJ_{n+2}\!\left(\frac{\wt_{T_{12}}}{2\omega_p}\right)\right],\\
\Omega_{02}^{(n)}&=\nb_{02}\left[\BJ_{n-1}\!\left(\frac{\wt_{T_{02}}}{2\omega_p}\right)-\BJ_{n+1}\!\left(\frac{\wt_{T_{02}}}{2\omega_p}\right)\right],
\end{align}
with
\begin{align}
\nu_{01}(t)&=\omega_p\phit_p\frac{\lambda_T(t)}{4\sqrt{\xi_T(t)}}\frac{E_{J_{T_1}}E_{J_{T_2}}}{E_{J_\eff}^2(t)}\left[\frac{E_{J_{T_1}}}{E_{J_{T_2}}}+\cos\phi_\ext(t)\right],\\
\nu_{12}(t)&=\omega_p\phit_p\frac{\Lambda_T(t)}{2\sqrt{2\xi_T(t)}}\frac{E_{J_{T_1}}E_{J_{T_2}}}{E_{J_\eff}^2(t)}\left[\frac{E_{J_{T_1}}}{E_{J_{T_2}}}+\cos\phi_\ext(t)\right],\\
\nu_{02}(t)&=\omega_p\phit_p\BJ_1(\phit_p)\frac{E_{J_{T_1}}E_{J_{T_2}}}{4E_{J_\eff}^2(t)}\upsilon(t).
\end{align}
\raggedbottom

\end{document}